\documentclass[preprint,aps,jcp,onecolumn,superscriptaddress]{revtex4-2}

\usepackage[paperheight=11in,paperwidth=8.25in,margin=0.67in]{geometry}

\usepackage[T1]{fontenc} 
\usepackage[version=3]{mhchem} 
\usepackage{mathpazo}

\usepackage{amsfonts}
\usepackage{amsmath}
\usepackage{booktabs}
\usepackage{tikz}
\usepackage{threeparttable}
\usepackage{multirow}
\usepackage{appendix}
\usepackage{amssymb}
\usepackage{graphicx}
\usepackage{subfigure}
\usepackage{enumerate}
\usepackage{colortbl}
\usepackage{framed}
\usepackage{verbatim}
\usepackage{amsbsy}
\usepackage{bm}

\usepackage[normalem]{ulem}
\usepackage{float}
\usepackage[linesnumbered,boxed]{algorithm2e}
\usepackage{algorithm2e}

\usepackage{simplewick} 
\usepackage{bbm}

\newcommand{\ii}{\mathbbm{i}}
\newcommand{\sgn}{ {\rm{sgn}} }

\newfloat{Algorithm}{htbp}{alg}
\floatname{Algorithm}{Algorithm}

\newcounter{exe}[figure]
\newcommand{\iexe}{\refstepcounter{exe}\the\value{exe}:}

\begin{document}

\title{Response to ``Response to `Comment on Theoretical examination of QED Hamiltonian in relativistic molecular orbital theory''' [J. Chem. Phys. 160, 187102 (2024)]}

\author{Wenjian Liu}\email{liuwj@sdu.edu.cn}
\affiliation{Qingdao Institute for Theoretical and Computational Sciences,
School of Chemistry and Chemical Engineering, Shandong University, Qingdao, Shandong 266237, P. R. China}

\maketitle

The Response of Inoue and coworkers\cite{RespCommentQED} to my Comment\cite{CommentQED} on their Paper I\cite{JapQED} clarifies some points,
but also raises some new issues that require further clarifications.

To begin with, it should be noted that any second quantization must
start with a definition of the fermion field operator $\hat{\phi}$ in terms of
a vacuum and corresponding annihilation and creation operators, such that the resulting
Hamiltonian takes the whatsoever vacuum as its ground state of zero energy. For instance,
if we start with
\begin{align}
\hat{\phi}&=a_p\psi_p,\quad
a_p|\mathrm{vac}\rangle=0,\quad p \in \mbox{PES, NES}, \label{vac0}
\end{align}
we will have the following Hamiltonian (in the notation of the Comment\cite{CommentQED})
\begin{align}
\mathcal{H}&=D_p^qa^p_q +\frac{1}{2}g_{pr}^{qs}a^{pr}_{qs},\quad p, q, r, s \in \mbox{PES, NES},\label{Hbase}
\end{align}
which is already normal ordered with respect to $|\mathrm{vac}\rangle$. However, this Hamiltonian does not distinguish the empty from the
filled Dirac picture. Since the empty Dirac picture must be abandoned (for it implies that no atom would be stable),
we ought to incorporate the field Dirac picture, by introducing a reference $|0;\tilde{N}\rangle$ ($=\Pi_{\tilde{i}}^{\tilde{N}} a^{\tilde{i}}|0;\tilde{0}\rangle$)
built up with zero positive energy states (PES) and $\tilde{N}$
($\rightarrow\infty$) negative energy states (NES). The Hamiltonian \eqref{Hbase} must then be normal ordered with respect to
$|0;\tilde{N}\rangle$, 
so as to obtain the physical Hamiltonian $H_n^{\mathrm{QED(MO)}}$\cite{eQED,PhysRep}
\begin{align}
\mathcal{H}&=H_n^{\mathrm{QED(MO)}} + C_n^{\mathrm{QED(MO)}},\label{H2}\\
H_n^{\mathrm{QED(MO)}}&=H^{\mathrm{FS(MO)}}_n + Q_p^q\{a^p_q\}_n,\quad p, q, r, s \in \mbox{PES, NES},\label{HnH}\\
H^{\mathrm{FS(MO)}}_n&=D_p^q\{a^p_q\}_n +\frac{1}{2}g_{pr}^{qs}\{a^{pr}_{qs}\}_n,\label{KZFS}\\
Q_p^q&=\bar{g}^{qs}_{pr} \acontraction[0.5ex]{}{a^r}{}{a_s}a^ra_s,\quad \bar{g}_{pr}^{qs}=g^{qs}_{pr}-g^{sq}_{pr},\label{Qdef}\\
C_n^{\mathrm{QED(MO)}}&=\langle 0;\tilde{N}|H|0;\tilde{N}\rangle\nonumber \\
&=D_p^{q}\acontraction[0.5ex]{}{a^p}{}{a_q}a^pa_q+\frac{1}{2}\bar{g}_{pr}^{qs}\acontraction[0.5ex]{}{a^p}{}{a_q}a^pa_q \acontraction[0.5ex]{}{a^r}{}{a_s}a^ra_s,\label{HnC}
\end{align}
where the contraction can be one of the three cases
\begin{align}
\acontraction[0.5ex]{}{a^p}{}{a_q}a^pa_q&=
\begin{cases}
0 \quad \mbox{ (constantly null contraction (CNC))};\\
\langle 0,\tilde{N}|a^p_q|0,\tilde{N}\rangle \quad \mbox{ (conventional contraction (CC))};\\
\frac{1}{2}\langle 0,\tilde{N}|[a^p, a_q]|0,\tilde{N}\rangle \quad \mbox{ (charge-conjugated contraction (CCC))}.\label{CCC}
\end{cases}
\end{align}
Note that the CC is unphysical\cite{LiuQED2020} and is included here only for later use.
By making use\cite{RespCommentQED} of the elementary anticommutation relation (ACR)
\begin{subequations}\label{Commut}
\begin{equation}
a^p a^q + a^q a^p=0,
\end{equation}
\begin{equation}
a_pa_q + a_qa_p=0,
\end{equation}
\begin{equation}
a^p a_q + a_q a^p=\delta_{pq},\label{Fermi}
\end{equation}
\end{subequations}
the (generic) Hamiltonian\cite{eQED,PhysRep} $H_n^{\mathrm{QED(MO)}}$ \eqref{HnH} was rewritten in Paper I\cite{JapQED}
in terms of unordered operators
\begin{subequations}\label{unnormalOP}
\begin{equation}
O_p^q\{a^p_q\}_n=O_p^qa^p_q-O_{\tilde{i}}^{\tilde{i}},\quad O= D, Q,
\end{equation}
\begin{equation}
\frac{1}{2}g_{pr}^{qs}\{a^{pr}_{qs}\}_n=\frac{1}{2}g_{pr}^{qs} a^{pr}_{qs}-\bar{g}_{p\tilde{j}}^{q\tilde{j}}a^p_q+\frac{1}{2}\bar{g}_{\tilde{i}\tilde{j}}^{\tilde{i}\tilde{j}},
\end{equation}
\end{subequations}
so as to obtain
\begin{align}
H^{\mathrm{QED(MO)}}_n&=\mathcal{H}+(Q_p^q-\bar{g}_{p\tilde{j}}^{q\tilde{j}})a^p_q + \bar{C}_n^{\mathrm{QED(MO)}},\label{nuQED}\\
\bar{C}_n^{\mathrm{QED(MO)}}&=-(Q+D)_{\tilde{i}}^{\tilde{i}}+\frac{1}{2}\bar{g}_{\tilde{i}\tilde{j}}^{\tilde{i}\tilde{j}},
\end{align}
where
\begin{align}
Q_p^q&=
\begin{cases}
0 \quad \mbox{ (CNC)},\\
\bar{g}_{p\tilde{j}}^{q\tilde{j}} \quad \mbox{ (CC)},\\
-\frac{1}{2}\bar{g}_{ps}^{qs}\sgn({\epsilon_s}) \quad \mbox{ (CCC)}.\label{Qpot}
\end{cases}
\end{align}
A close inspection reveals that the Hamiltonian $\mathcal{H}$ in Eq. \eqref{nuQED}
must adopt the CC [which is in line with Eq. \eqref{Commut}] in order to be consistent. Therefore,
Eq. \eqref{nuQED} had better be rewritten as
\begin{align}
H^{\mathrm{QED(MO)}}_{n(u)}&=\mathcal{H}_{\mathrm{CC}}+(Q_p^q-\bar{g}_{p\tilde{j}}^{q\tilde{j}})a^p_q + \bar{C}_n^{\mathrm{QED(MO)}}\label{nuQEDCC}
\end{align}
in order to be precise. As such, Eq. \eqref{nuQED}/\eqref{nuQEDCC} is indeed a hybrid\cite{CommentQED} of ``two pictures'',
CNC/CCC vs. CC. Note in passing that
Eq. \eqref{nuQED} can also be understood as
a reexpression of $H^{\mathrm{QED(MO-X)}}_n$ (X = CNC, CC, CCC) in terms of $H^{\mathrm{QED(MO-CC)}}_n$, i.e.,
\begin{align}
H^{\mathrm{QED(MO-X)}}_n=H^{\mathrm{QED(MO-CC)}}_n+(Q-\bar{g}_{p\tilde{j}}^{q\tilde{j}})\{a^p_q\}_n,\label{junk}
\end{align}
which is then converted to the formally unordered form \eqref{nuQEDCC} (indicated by the subscript $n(u)$).

Since quantities like $D_{\tilde{i}}^{\tilde{i}}$, $\bar{g}_{\tilde{i}\tilde{j}}^{\tilde{i}\tilde{j}}$, and especially $\bar{g}_{p\tilde{j}}^{q\tilde{j}}a^p_q$ are all
divergent (due to the existence of infinitely many NES $\{\tilde{i},\tilde{j}\}$), they cannot be evaluated numerically.
Rather, they must be canceled out analytically when calculating the physical energy of an $N$-electron system according to
\begin{align}
E&=\langle \Psi(N;\tilde{N})|H^{\mathrm{QED(MO)}}_{n(u)}|\Psi(N;\tilde{N})\rangle-\langle\Psi(0;\tilde{N})|H^{\mathrm{QED(MO)}}_{n(u)}|\Psi(0;\tilde{N})\rangle.\label{PhysEnCC}
\end{align}
Here, the (normalized) wave function $|\Psi(0;\tilde{N})\rangle$ represents a polarizable vacuum, with $|0;\tilde{N}\rangle$ as its zeroth order\cite{PhysRep}.
As shown by Inoue and coworkers\cite{RespCommentQED}, the divergent terms appearing in Eqs. (13), (16),
 (33), (81), and (99) in Paper I\cite{JapQED} do disappear after working out
the matrix elements explicitly. Therefore, the criticisms on these equations made in the Comment\cite{CommentQED} should be eliminated.
Still, however, this twisted form \eqref{nuQEDCC} of the Hamiltonians is unpleasant, for it looks like
that the divergent terms, which are not present in the elegant expressions of $H^{\mathrm{QED(MO-CNC)}}_n$/$H^{\mathrm{FS(MO)}}_n$ \eqref{KZFS} and $H^{\mathrm{QED(MO-CCC)}}_n$ \eqref{HnH},
are first extracted out deliberately and then canceled out carefully. In our opinion, it can be
viewed at most as an intermediate quantity (instead of a proper form of the Hamiltonians)
for the evaluation of the energy \eqref{PhysEnCC}. Actually, it is not needed even for the calculation of the energy, for which
can actually be calculated as\cite{PCCPNES,PhysRep}
\begin{align}
E&=\langle \Psi(N;\tilde{N})|H^{\mathrm{QED(MO)}}_n|\Psi(N;\tilde{N})\rangle-\langle\Psi(0;\tilde{N})|H^{\mathrm{QED(MO)}}_n|\Psi(0;\tilde{N})\rangle\label{PhysEn}\\
&=\langle \Psi(N;\tilde{N})|\mathcal{H}-C^{\mathrm{QED(MO)}}_n|\Psi(N;\tilde{N})\rangle-\langle\Psi(0;\tilde{N})|\mathcal{H}-C^{\mathrm{QED(MO)}}_n|\Psi(0;\tilde{N})\rangle\label{PhysEn1}\\
&=\langle \Psi(N;\tilde{N})|\mathcal{H}|\Psi(N;\tilde{N})\rangle-\langle\Psi(0;\tilde{N})|\mathcal{H}|\Psi(0;\tilde{N})\rangle.\label{PhysEnu}
\end{align}
It is just that the same CNC/CCC contraction should be followed when going from Eq. \eqref{PhysEn} to \eqref{PhysEnu}. To the best of our knowledge,
this ansatz was first proposed in Ref. \citenum{PCCPNES} and presented more clearly
in Ref. \citenum{PhysRep}. It can readily checked that the CC even does not yield the correct
first order energy $E^{(1)}_{\mathrm{CC}}$, viz.,
\begin{align}
E^{(1)}_{\mathrm{CC}}&=\langle N;\tilde{N}|H^{\mathrm{QED(MO-CC)}}_n|N;\tilde{N}\rangle\nonumber\\
&=D_i^i+\frac{1}{2}\bar{g}_{ij}^{ij}+\bar{g}_{i\tilde{j}}^{i\tilde{j}}\rightarrow +\infty, \quad i, j\in\mbox{ PES}, \tilde{j}\in\mbox{ NES},
\end{align}
and should hence be rejected from the outset.

At variance with Eq. \eqref{PhysEn}/\eqref{PhysEnu}, the energy $E$ of an $N$-electron system can also be calculated either as\cite{eQED}
\begin{align}
E&=\langle N;\tilde{N}|H^{\mathrm{QED(MO)}}_n|\bar{\Psi}(N;\tilde{N})\rangle,\quad |\bar{\Psi}(N;\tilde{N})\rangle=
\Omega_n|N;\tilde{N}\rangle, \label{MBPT}
\end{align}
or as\cite{PhysRep}
\begin{align}
E&=      \langle\bar{\Psi}(N;\tilde{N})| H^{\mathrm{QED(MO)}}_n|\bar{\Psi}(N;\tilde{N})\rangle \label{ExpectEn}\\
 &\equiv \langle\bar{\Psi}(N;\tilde{N})| \mathcal{H}  |\bar{\Psi}(N;\tilde{N})\rangle-C^{\mathrm{QED(MO)}}_n. \label{ExpectEnu}
\end{align}
The two differ only in normalization (intermediate vs. unitary normalization).
Again, the same CNC/CCC contraction should be followed when going from Eq. \eqref{ExpectEn} to \eqref{ExpectEnu}.
A subtle but important point lies in that the wave function $|\bar{\Psi}(N;\tilde{N})\rangle$
is here generated by the action of a normal-ordered wave operator $\Omega_n$ on the non-interacting reference $|N;\tilde{N}\rangle$.
As such, a polarizable vacuum is not needed here (for detailed derivations, see Section IIB.2 and Appendix A in Ref. \citenum{eQED}).
The so-calculated energy is identical to that obtained by Eq. \eqref{PhysEnu}/\eqref{PhysEnCC} and is
in termwise agreement with that by the S-matrix formulation of QED\cite{eQED}.
With this more transparent presentation (as compared to Eq. (27) in Ref. \citenum{PhysRep} or Eqs. (48) and (49) in the Comment\cite{CommentQED}),
it should be clear that
the remark (on Eq. \eqref{ExpectEnu}) that ''the terms subtracted from the referenced Hamiltonians
are single Slater determinants and cannot remove total energy divergence caused by the generalized
electron correlation''\cite{JapQED} does not hold true.

As far as the nonrelativistic limit of $H^{\mathrm{QED(MO-CCC)}}_n$ \eqref{HnH} is concerned, Inoue and coworkers
obviously did not recognize the physical meaning of the $Q$-potential \eqref{Qdef}/\eqref{Qpot} arising from the CCC \eqref{CCC}.
As emphasized from its advent\cite{eQED} (see also Ref. \citenum{LiuQED2020} for a thorough discussion),
the direct and exchange parts of $Q$ \eqref{Qpot} represent the vacuum polarization (VP) and electron self-energy (ESE), respectively
(NB: the frequency-dependent transverse-photon interaction has to be further added to the exchange part to make the ESE complete\cite{PhysRep}).
It is then textbook knowledge that both the VP and ESE have to be regularized/renormalized before they can actually be
used (see Ref. \citenum{VPregularizationPRA2003} for numerical demonstrations).
After regularization/renomalization, the $Q$-potential is at most of $\mathcal{O}(Z^3\alpha^3)$ and therefore vanishes in the nonrelativistic limit.
That is, the two terms in Eq. (10) in the Response\cite{RespCommentQED} are divergent individually
but tend to cancel each other, and can hence be renomalized away. This is an essential feature of QED.
Speaking of ``QED Hamiltonian'' in the absence of the VP-ESE (\emph{genuine QED effect}\cite{eQED}) is hardly meaningful.
What were reported in Paper I\cite{JapQED} are actually relativistic configuration interaction and
many-body perturbation theories of composite systems of $N$ electrons and $\tilde{M}$ (real) positrons governed by
the instantaneous Coulomb interaction, instead of
the initially anticipated\cite{CommentQED} contribution of the NES (virtual positrons) to
the correlation of $N$ electrons, a kind of \emph{derived QED effect}\cite{eQED}. For instance,
Eqs. (B1), (B4), and B(6) in the Response \cite{RespCommentQED} all go back to the corresponding no-pair expressions
for $N$ electrons by setting the number of positrons to zero. It should be restated that Eq. (95) in Paper I\cite{JapQED}
misses by construction the one-body term $E_1^{(2)}$ in Eq. (52) in the Comment\cite{CommentQED}.
Likewise, their ``QED-based'' Dirac-Hartree-Fock (DHF) theory (see Section IIC.1 in Paper I) is also merely
a relativistic but non-QED mean-field theory\cite{DyallPositron} of $N$ electrons and $\tilde{M}$ positrons.
The genuine QED mean-field theories for electrons only and for both electrons and positrons
were presented in Ref. \citenum{PhysRep} (see Eq. (54)/(81) or (85) therein) and Ref. \cite{LiuQED2020,eQEDBook2017}, respectively.
In such genuine QED mean-field theories, the PES and NES, whether occupied or not,  are all coupled, thereby fundamentally different
from the relativistic but non-QED counterparts.

Inoue and coworkers also made an interesting observation\cite{RespCommentQED}: in contrast to the CC,
both the CNC and CCC do not satisfy the basic ACR of fermion operators\eqref{Commut}.
At first glance, this might lead to that the physically incorrect CC is good, but the physically correct CCC/CNC is bad!
To resolve this apparent self-contradiction, they separated the CNC/CCC from the
ACR (or equivalently the CC), and took the former only as a formal-theoretical process for
constructing the Hamiltonians. This route is certainly not incorrect. However, it is not a proper interpretation.
In sharp contrast, the introduction of the CCC\cite{eQED}
is firmly founded, not merely a formal (mathematical) step. As scrutinized in depth in Ref. \citenum{LiuQED2020},
the introduction of a filled Dirac sea of electrons (which is not part of the Dirac equation itself) is a must but
not yet complete. By charge conjugation symmetry (which is indeed a property of the Dirac equation),
there exists also a filled Dirac sea of positrons (which is again not part of the Dirac equation itself).
The two seas are coexistent and equivalent, and should hence be averaged with an equal weight of one half,
thereby leading naturally to the CCC\cite{eQED}. Just like that the ACR is dictated by fermi statistics
of nonrelativistic or relativistic fermions, the CCC is dictated by charge conjugation symmetry
of relativistic fermions. Both are fundamental laws of physics that
have to be imposed from the outside of quantum mechanical equations, so as to render the latter physically correct
(recalling that the Schr\"odinger equation can also describe bosons, provided that boson statistics is mposed).
Note in passing that the CCC is just the time-independent analog
of the symmetric-in-time, equal-time contraction (ETC) of time-dependent fermion operators\cite{Schwinger1951}, which
is embodied automatically in the Feynman fermion propagator\cite{IJQCeQED}, again not merely a formal ingredient.
The correct four-current for electrons (and positrons), and hence the VP-ESE represented by the $Q$-potential \eqref{Qdef}/\eqref{Qpot},
can only be obtained by the CCC/ETC (see Eqs. (36)--(58) in Ref. \citenum{LiuQED2020}).
Yet, it should be kept in mind that
the CCC is to be applied only when normal ordering with respect to the NES (virtual positrons).
In contrast, the CC (or equivalently the ACR) should still be applied when further normal ordering with respect to the occupied PES\cite{PhysRep},
where charge conjugation is irrelevant. It is in this sense that the $Q$-potential arising from the CCC (zero from the CNC)
should be interpreted as an integral part of the Hamiltonian, as implied in Eqs. \eqref{nuQED}, \eqref{nuQEDCC}, and \eqref{junk}.
In other words, being imposed from the outset, the CCC $\acontraction[0.5ex]{}{a^p}{}{a_q}a^pa_q$ over the vacuum $|0;\tilde{N}\rangle$
should be viewed as an integral part of the normal ordering process.
It can also be shown that the CCC arises naturally from the averaging of
the Hamiltonians for electrons and positrons moving in the same external field (which is a must, as shown in Appendix \ref{CCC-H}).

Since the Hamiltonian $H^{\mathrm{QED(MO-CNC)}}$ misses by construction the \emph{genuine QED effect}\cite{eQED}, it had better go back to its original name,
``no-photon (non-QED) Fock space Hamiltonian'', as advocated by Kutzelingg more than a decade ago\cite{KutzFS}.

In summary, the major conclusion in the Comment\cite{CommentQED} remains unaltered: it is the CCC-based
Hamiltonian $H^{\mathrm{QED(MO-CCC)}}$, instead of the CNC-based Hamiltonian
$H^{\mathrm{QED(MO-CNC)}}$/$H^{\mathrm{FS(MO)}}$, that is a genuine QED Hamiltonian, and should hence
be recommended as the basis of the emerging field of ``molecular QED''\cite{PhysRep}.

\section*{Acknowledgement}
This work was supported by the National Natural Science Foundation of China (Grant No. 22373057) and
Mount Tai Scholar Climbing Project of Shandong Province.

\section*{Data Availability Statement}
The data that supports the findings of this study is available within the article.

\appendix
\section{Hamiltonian quantized with charge conjugated fields}\label{CCC-H}
The charge conjugation is defined as
\begin{equation}
\hat{C}=\mathbf{C}_0\hat{K}_0,\quad \hat{C}^\dag=\hat{C}^{-1}=\hat{C},
\quad \mathbf{C}_0=-\ii \alpha_y\beta=\begin{pmatrix}\mathbf{0}_2&i\sigma_y\\
-i\sigma_y&\mathbf{0}_2\end{pmatrix}=\mathbf{C}_0^\dag=\mathbf{C}_0^{-1},\label{CCsymm}
\end{equation}
which transforms the one-electron Dirac equation
\begin{align}
D\psi_p&=\epsilon_p\psi_p,\\
D&=D_0+q\phi_{\mathrm{ext}}(\boldsymbol{r}),\quad q=-1,\\
D_0&=c\boldsymbol{\alpha}\cdot\boldsymbol{p}+\beta mc^2
\end{align}
to that of a positron
\begin{align}
D^C \psi^C_p &=\epsilon_p^C \psi^C_p,\label{PositronDEQ}\\
D^C&=-\hat{C}D\hat{C}^{-1}
=-\mathbf{C}_0^\dag D^* \mathbf{C}_0=D_0-q\phi_{\mathrm{ext}}(\boldsymbol{r}),\label{hDcop} \\
\psi_p^C&=\hat{C}\psi_p=\mathbf{C}_0\psi^*_p,\quad \epsilon_p^C=-\epsilon_p.
\end{align}
Likewise, the four-component Dirac field operator $\hat{\phi}(\boldsymbol{r})$ (in the particle-hole picture)
\begin{align}
\hat{\phi}_{\sigma}(\boldsymbol{r})&=b_p\psi_{p\sigma}(\boldsymbol{r})+b^{\tilde{p}}\psi_{\tilde{p}\sigma}(\boldsymbol{r}),
\quad p\in\mbox{ PES}, \quad \tilde{p}\in\mbox{ NES},\quad \sigma\in[1,4]\label{ElectronField}
\end{align}
will be transformed by charge conjugation to
\begin{align}
\hat{\phi}^C(\boldsymbol{r})&=\mathbf{C}_0 \hat{\phi}^{\dag T}(\boldsymbol{r}) \\
&=b_{\tilde{p}}\psi^C_{\tilde{p}}(\boldsymbol{r})+b^p\psi_p^C(\boldsymbol{r}),\label{PositronField}
\end{align}
where the first and second terms annihilate a positive-energy positron (NB: $\epsilon_{\tilde{p}}^C>0$) and create a positive-energy electron, respectively,
in accordance with the fact that the first and second terms of $\hat{\phi}(\boldsymbol{r})$ \eqref{ElectronField}
annihilate a positive-energy electron and create a positive-energy positron, respectively. That is,
$\hat{\phi}^C(\boldsymbol{r})$ \eqref{PositronField} would be the starting quantized Dirac field, had we lived in the world of antiparticles.

Both $\hat{\phi}(\boldsymbol{r})$ \eqref{ElectronField} and
$\hat{\phi}^C(\boldsymbol{r})$ \eqref{PositronField} are associated with the vacuum $|\mathrm{vac}\rangle=|0;\tilde{0}\rangle$.
To expedite algebraic manipulations, we can rewrite them as\cite{LiuQED2020}
\begin{align}
\hat{\phi}(\boldsymbol{r})&=a_p\psi_p(\boldsymbol{r})+a_{\tilde{p}}\psi_{\tilde{p}}(\boldsymbol{r}),
\quad \epsilon_p>0, \quad \epsilon_{\tilde{p}}<0,\label{ElectronFielda}\\
\hat{\phi}^C(\boldsymbol{r}) &=a_{\tilde{p}}\psi^C_{\tilde{p}}(\boldsymbol{r})+a_p\psi_p^C(\boldsymbol{r}),
\quad \epsilon_{\tilde{p}}^C>0,\quad \epsilon_p^C<0,\label{PositronFielda}
\end{align}
by taking $|0;\tilde{N}\rangle$ (Dirac sea of electrons)
and $|0_{e^+};\tilde{N}_{e^+}\rangle$ (Dirac sea of positrons)
as the vacua, respectively.

With the above background, we can calculate
\begin{align}
h^C&=\int \hat{\phi}^{C\dag} D^C \hat{\phi}^C d\tau=\int (\mathbf{C}_0 \hat{\phi}^{\dag T})^\dag D^C (\mathbf{C}_0 \hat{\phi}^{\dag T}) d\tau
=\int \hat{\phi}^{T} (\mathbf{C}_0^\dag D^C \mathbf{C}_0) \hat{\phi}^{\dag T}d\tau\nonumber\\
&=-\int \hat{\phi}^{T} D^* \hat{\phi}^{\dag T}d\tau=-\int \hat{\phi}^{T} D^T \hat{\phi}^{\dag T}d\tau\nonumber\\
&=-\int \hat{\phi}_{\rho} D_{\sigma\rho}\hat{\phi}^{\dag}_{\sigma}=-\int D \hat{\phi}\hat{\phi}^{\dag}d\tau\nonumber\\
&=\int \{\hat{\phi}^{\dag}D \hat{\phi}\}_nd\tau-\int \langle\mathrm{vac}|D\hat{\phi}\hat{\phi}^\dag|\mathrm{vac}\rangle  d\tau\label{hCpositron}\\
&=h_n - D_p^q \langle 0;\tilde{N}|a_q a^p|0;\tilde{N}\rangle,\quad p,q\in\mbox{ PES, NES},\label{hCpositrona}\\
h_n&=D_p^q\{a^p_q\}_n,\label{hnOp}
\end{align}
which is to be compared to the usual second-quantized Dirac operator
\begin{align}
\tilde{h}&=\int \hat{\phi}^{\dag} D \hat{\phi} d\tau=\int \{\hat{\phi}^{\dag}D \hat{\phi}\}_n d\tau
+ \int \langle\mathrm{vac}|\hat{\phi}^\dag D\hat{\phi}|\mathrm{vac}\rangle d\tau\label{helectron}\\
&=h_n + D_p^q \langle 0;\tilde{N}|a^p a_q|0;\tilde{N}\rangle,\quad p,q\in\mbox{ PES, NES}.\label{helectrona}
\end{align}
Since charge conjugation is an inherent symmetry, $\tilde{h}$ \eqref{helectron}/\eqref{helectrona} and
$h^C$ \eqref{hCpositron}/\eqref{hCpositrona} should be averaged with an equal weight, leading to
\begin{align}
h&= \frac{1}{2}(\tilde{h}+h^C)
=\int \{\hat{\phi}^{\dag}D \hat{\phi}\}_n d\tau +\frac{1}{2}\int \langle\mathrm{vac}|[\hat{\phi}^\dag, D\hat{\phi}]|\mathrm{vac}\rangle d\tau\\
&=h_n + C_{n1},\\
C_{n1}&= D_p^q \acontraction[0.5ex]{}{a^p}{}{a_q}a^pa_q, \quad \acontraction[0.5ex]{}{a^p}{}{a_q}a^pa_q=\frac{1}{2}\langle 0;\tilde{N}|a^p, a_q|0;\tilde{N}\rangle
=-\frac{1}{2}\delta_{pq}\sgn(\epsilon_p),
\end{align}
where $\acontraction[0.5ex]{}{a^p}{}{a_q}a^pa_q$ is just the CCC \eqref{CCC} and arises here naturally from the averaging process.

The two-body operator can be calculated in the same way, viz.,
\begin{align}
G^C &= \frac{1}{2}\int\int \hat{\phi}^{C\dag}(1) \hat{\phi}^{C\dag}(2) V(1,2) \hat{\phi}^{C}(2)\hat{\phi}^{C}(1) d\tau_1 d\tau_2\nonumber\\
&=\frac{1}{2}\int\int \hat{\phi}^T(1)\hat{\phi}^T(2)[\mathbf{C}_0^\dag(1)\mathbf{C}_0^\dag(2)V(1,2)\mathbf{C}_0(2)\mathbf{C}_0(1)]
\hat{\phi}^{\dag T}(2)\hat{\phi}^{\dag T}(1)\nonumber\\
&=\frac{1}{2}\int\int \hat{\phi}^T(1)\hat{\phi}^T(2)V^*(1,2)
\hat{\phi}^{\dag T}(2)\hat{\phi}^{\dag T}(1)\nonumber\\
&=\frac{1}{2}\int\int\hat{\phi}_{\rho}(1)\hat{\phi}_{\sigma}(2)[V(1,2)]_{\delta\rho,\gamma\sigma  }\hat{\phi}^\dag_{\gamma}(2)\hat{\phi}^\dag_{\delta}(1)\nonumber\\
&=\frac{1}{2}\int\int V(1,2)\hat{\phi}(1)\hat{\phi}(2)\hat{\phi}^\dag (2)\hat{\phi}^\dag (1)\\
&=\frac{1}{2}g_{pr}^{qs} a_qa_s a^r a^p\\
&=\frac{1}{2}g_{pr}^{qs}\{a^{pr}_{qs}\}_n -\bar{g}_{pr}^{qs}\langle 0;\tilde{N}|a_sa^r|0;\tilde{N}\rangle\{a^p_q\}_n
+\frac{1}{2}\bar{g}_{pr}^{qs}\langle 0;\tilde{N}|a_qa^p|0;\tilde{N}\rangle\langle 0;\tilde{N}|a_sa^r|0;\tilde{N}\rangle,\label{GCnew}
\end{align}
which should be compared to the usual two-body operator
\begin{align}
\tilde{G}&= \frac{1}{2}\int\int \hat{\phi}^{\dag}(1) \hat{\phi}^{\dag}(2) V(1,2) \hat{\phi}(2)\hat{\phi}(1) d\tau_1 d\tau_2\nonumber\\
&=\frac{1}{2}g_{pr}^{qs} a^{pr}_{qs}\\
&=\frac{1}{2}g_{pr}^{qs}\{a^{pr}_{qs}\}_n+\bar{g}_{pr}^{qs}\langle 0;\tilde{N}|a^ra_s|0;\tilde{N}\rangle\{a^p_q\}_n
+\frac{1}{2}\bar{g}_{pr}^{qs}\langle 0;\tilde{N}|a^pa_q|0;\tilde{N}\rangle\langle 0;\tilde{N}|a^ra_s|0;\tilde{N}\rangle. \label{Gnew}
\end{align}
The average of $\tilde{G}$ and $G^C$ leads to
\begin{align}
G&=\frac{1}{2}(\tilde{G}+G^C)=G_n+C_{n2},\\
G_n&=\frac{1}{2}g_{pr}^{qs}\{a^{pr}_{qs}\}_n + Q_p^q \{a^p_q\}_n,\quad Q_p^q =\bar{g}_{pr}^{qs} \acontraction[0.5ex]{}{a^r}{}{a_s}a^ra_s,
\quad \quad \acontraction[0.5ex]{}{a^r}{}{a_s}a^ra_s=\frac{1}{2}\langle 0;\tilde{N}|a^r, a_s|0;\tilde{N}\rangle,\label{GnOp}\\
C_{n2}&=\frac{1}{2}\bar{g}_{pr}^{qs} [\langle 0;\tilde{N}|a^pa_q|0;\tilde{N}\rangle\langle 0;\tilde{N}|a^ra_s|0;\tilde{N}\rangle
+  \langle 0;\tilde{N}|a_pa^q|0;\tilde{N}\rangle\langle 0;\tilde{N}|a_sa^r|0;\tilde{N}\rangle]. \label{Cn2}
\end{align}
It can be seen that the sum of $h_n$ \eqref{hnOp} and $G_n$ \eqref{GnOp} is just $H_n^{\mathrm{QED(MO)}}$ \eqref{HnH} along with the CCC. In particular,
in sharp contrast to the $Q$-potential in Eq. \eqref{GnOp},
the second term of $G^C$ \eqref{GCnew} or that of $\tilde{G}$ \eqref{Gnew} is divergent and cannot be rgularized/renormalized.
As such, the averaging of the Hamiltonian $\tilde{h}+\tilde{G}$ for electrons and the Hamiltonian $h^C+G^C$ for positrons (moving in the
same external field $\phi_{\mathrm{ext}}(\boldsymbol{r})$ as electrons)
is a must rather than merely a formal step. Note in passing that
the constant $C_{n2}$ \eqref{Cn2} is different from the second term of Eq. \eqref{HnC} (which is more symmetric  due to
the direct use of the CCC). However, the difference does doe not matter at all, for such constants will be renormalized away.

The above manipulation reveals that the genuine QED Hamiltonian $H_n^{\mathrm{QED(MO-CCC)}}$ \eqref{HnH}, especially the Q-potential in Eq. \eqref{GnOp}
[or Eq. \eqref{Qdef}/\eqref{Qpot}],
does stem from the symmetric treatment of electrons and positrons, as dictated by charge conjugation symmetry of relativistic fermions.
In particular, Eqs. \eqref{hCpositrona}, \eqref{helectrona}, \eqref{GCnew}, and \eqref{Gnew} all follow strictly the ACR \eqref{Commut}.

\clearpage
\newpage

\bibliographystyle{apsrev4-2}
\bibliography{BDFlib}

\end{document}